\documentclass{mem}
\usepackage{natbib}\usepackage{txfonts}\usepackage{balance}
\usepackage{graphicx}
\usepackage[a4paper]{hyperref}
\def\ltsima{$\; \buildrel < \over \sim \;$}
\def\lsim{\lower.5ex\hbox{\ltsima}}
\def\gtsima{$\; \buildrel > \over \sim \;$}
\def\gsim{\lower.5ex\hbox{\gtsima}}

\newcommand{\be}{\begin{equation}}
\newcommand{\en}{\end{equation}}

\def\kms  {\rm \ km \, s^{-1}}

\begin{document}
\def\teff{$T\rm_{eff }$}
\def\kms{$\mathrm {km s}^{-1}$}

\title{Transient neutron star X--ray binaries 
with Simbol-X
}

   \subtitle{}

\author{
S. Campana\inst{1} 
          }

  \offprints{S. Campana}

\institute{
Osservatorio astronomico di Brera, Via E. Bianchi 46,
I-23807 Merate (LC), Italy
\email{sergio.campana@brera.inaf.it}
}

\authorrunning{S. Campana}

\titlerunning{Transient neutron stars with Simbol-X}

\abstract{
We present a brief overview of test-bed observations on accreting 
neutron star binaries for the Simbol-X mission. We show that Simbol-X
will provide unique observations able to disclose the physical 
mechanisms responsible for their high energy emission.

}
\maketitle{}

\section{Introduction: state of the art}

Low mass X--ray binaries (LMXBs) are binary systems composed by a 
compact object (here we consider only neutron stars, NS) and a low mass 
companion. These systems are the brightest X--ray emitters in the Galaxy, 
mainly distributed in the central regions and in the galactic bulge.
These sources come in two different flavours. There are persistent 
sources which shine at a steady (but variable) luminosity in the 
$10^{36}-10^{38}$ erg s$^{-1}$ range and transient sources which 
alternate intervals (lasting weeks to months) during which they 
rival in brightness with persistent sources to longer (years to decades)
quiescent periods (characterized by luminosities $10^{32}-10^{33}$ 
erg s$^{-1}$).

Bright persistent sources are variable on short timescales, with 
variation by a factor of a few on a daily timescale. Historically, 
in an X--ray color-color diagram they trace out either a Z (Z sources) 
or a C pattern (atoll sources, see van der Klis 1995 for a review). 
Z sources are the brightest NS X-ray binaries, do not show (Type I) 
X--ray bursts and display strong quasi-periodic oscillations (QPOs). 
Atoll sources are dim by a factor of $\lsim 10$, show X--ray bursts and 
display less intense QPOs. Transient NS binaries when 
in outburst show properties very similar to atoll sources.

\begin{figure}[]
\resizebox{\hsize}{!}{\includegraphics[clip=true,angle=-90]{campana_f1.ps}}
\resizebox{\hsize}{!}{\includegraphics[clip=true,angle=-90]{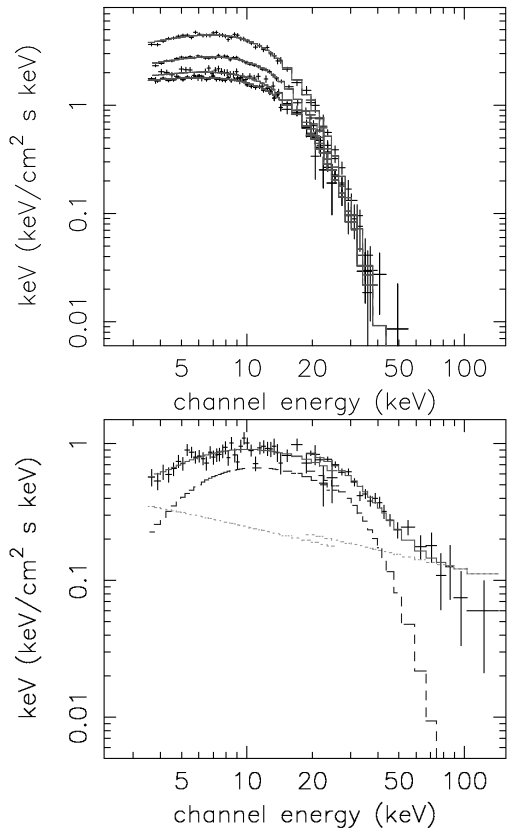}}
\caption{\footnotesize
Data and models of the soft spectra (top) and the hard spectrum (bottom) of
4U 1820--30 as observed by INTEGRAL during 2003-2005. The spectra were 
fitted by thermal Comptonization models without and with a power
law component, respectively (taken from Tarana et al. 2007).
}
\label{1820}
\end{figure}

The overall spectra of Z sources are soft (Barret \& Olive
2002, and references therein) and can be described by the sum
of a cool ($\sim 1$ keV) black body and a Comptonized emission
from an electron plasma (corona) of a few keV. Instead,
atoll sources undergo strong spectral changes: when
bright, they can have soft spectra (similar to Z sources) but
they switch to low/hard spectra at low luminosities (Barret 
\& Vedrenne 1994). Observationally, the soft and hard spectra 
of atoll sources are very different in the hard X--ray energy band:
the ratio between the hard (13--80 keV) to the soft luminosity 
(2--10 keV) may increase by a factor of $\sim 10$ (van Paradijs \& 
van der Klis 1994, see Fig. 1).
Further studies showed that there exists a limiting luminosity 
at which this change occurs, around a level of $\sim 4\times 10^{36}$ 
erg s$^{-1}$, and this spectral change is also common to black hole 
binaries (Maccarone 2003). Observationally, the Componized emission
changes from a few keV in the soft state to a few tens of keV 
in the hard state. 

Hard tails were historically observed also in Z sources, but 
only sporadically. The first detection was in the spectrum of Sco X-1
dominating the above 40 keV (Peterson \& Jacobson 1966). 
More recently the presence of a variable hard tail in Sco X-1 was 
confirmed by OSSE and RXTE observations (Strickman \& Barret 2000; 
D'Amico et al. 2001). BeppoSAX and INTEGRAL observations led to much 
progresses in this field. A hard component has been observed in several 
Z sources (see Di Salvo \& Stella 2002; Paizis et al. 2006) indicating 
that this is a common feature. This hard component, fitted by a power 
law with photon index in the range 1.9­-3.3, contributes up to $10\%$ 
to the source bolometric luminosity. 

A further step in the understanding the LMXB hard tails is provided 
by radio studies. A coupling between (hard) X--rays and
radio properties of NS LMXBs has been proven to exist (Migliari \& 
Fender 2006; Migliari et al. 2007). This can be seen in Fig. 2, showing 
the coupling between the radio jet emission and the position in the X--ray 
hardness-intensity diagram. In addition, a positive correlation between 
the radio flux density and the X--ray flux in the hard-tail power law component
has been found. These observations were interpreted as evidence for the
formation of a radio jet associated with the Flaring Branch-to-Normal Branch
X--ray state transition in the Z pattern (Migliari et al. 2007).

\begin{figure*}[]
\resizebox{\hsize}{!}{\includegraphics[clip=true]{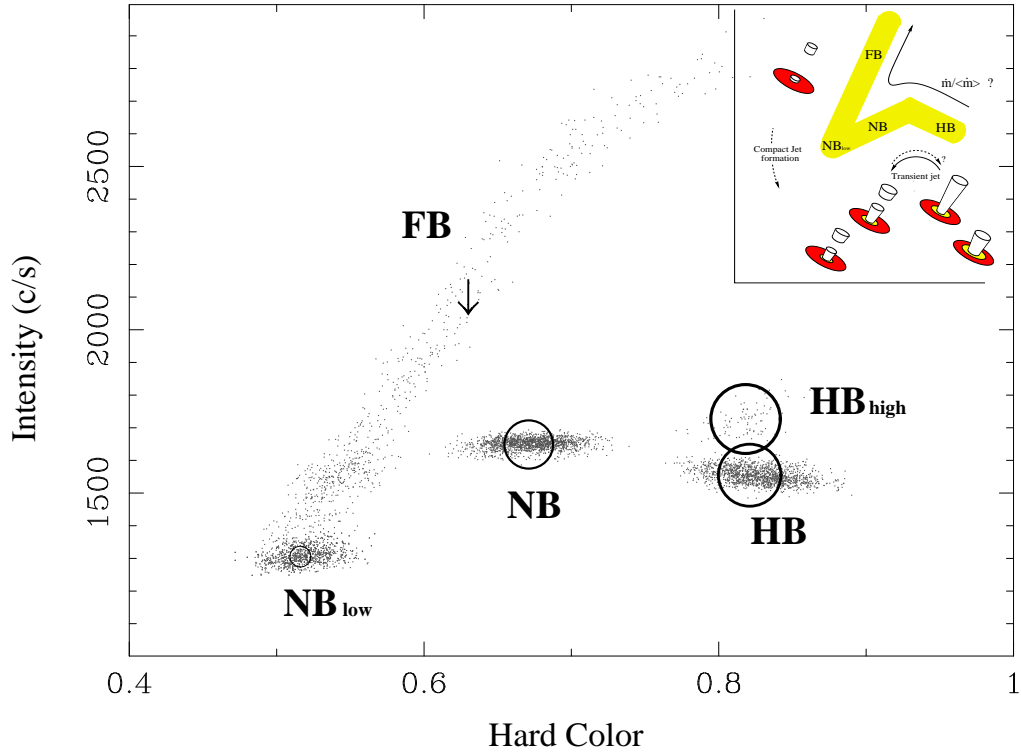}}
\caption{\footnotesize
Hardness intensity diagram of GX~17+2 (main panel), with a sketch of the
jet/X-ray state coupling (top-right panel) to include the formation of 
a compact jet at the FB-to-NB transition. In the main panel, the gray 
dots represent the HID of 16~s of observation. The open circles indicate
the mean radio flux density of the source observed in the different
branches: the bigger the circle, the higher the observed radio
flux density. The arrow in the FB indicates an upper limit on the radio
flux density. The radio emission is strongest in the HB$_{high}$ and, 
following the HID track, decreases towards the NB$_{low}$, until it is not
detectable anymore in the FB (from Migliari et al. 2007).
}
\label{gx17}
\end{figure*}

A close parallel between NS and black hole transients exists based 
on the connection between radio and hard X--ray emission. Black hole 
transients are better studied with a detection of A0620-­00 at an 
X--ray luminosities as low as $10^{-8.5}$ times the Eddington limit 
(Gallo et al. 2006). These observations favour a model for quiescence 
in which a radiatively inefficient outflow accounts for a sizable fraction
of the missing energy. This has not proven to be true in NS transient yet.
NS transients in quiescence are characterized by X--ray spectra made by 
a soft component (interpreted as emission coming from the cooling
of the neutron star atmosphere) and a power law tail (not present in all
systems and accounting for up to $50\%$ of the flux) of unknown origin
(e.g. Campana 2001). In addition, in some well studied sources this tail
has been shown to undergo substantial variations of unknown origin (Campana
\& Stella 2003). 

In the last few years it has become clear that the 
quiescent spectra of NS transients containing an accreting ms X--ray 
pulsar during outburst (e.g. SAX J1808.4--3658 and similar sources) 
lack of the soft component, being a dimmer ($\sim 10$) and showing only
a power law tail (Campana et al. 2004a; Wijnands et al. 2005; Heinke et 
al. 2007).
The physical nature of these tails is basically unknown (see Campana
et al. 1998 for some possibilities).

\section{Simbol-X on the scene}
 
Simbol-X will produce a revolution in the hard X--ray research 
field. Simbol-X will disclose the faint population of hard 
X--ray sources as the Einstein satellite did in the soft energy band, 
following the Uhuru satellite.
The main characteristics of Simbol-X are:

\begin{itemize}

\item the broad band capabilities with an effective energy 
range of 0.5--80 keV and possibly up to 100 keV;

\item the very large effective area at low (rivaling with XMM-Newton) 
and high energies;

\item a good timing capabilities without problems of pile-up
and/or storage problems, allowing for long observations of $\sim 0.5$ 
Crab sources; 

\item a very low background allowing for the detection and study of very dim
sources.

\end{itemize} 

\section{Simbol-X and compact objects}

In order to assess the potentialities of the Simbol-X mission we
consider here a few test cases.

\subsection{Study of hard tails in persistent neutron star sources}

We consider first an atoll source with an 0.5--20 keV X--ray flux 
of $10^{-9}$ erg cm$^{-2}$ s$^{-1}$ (i.e. about 60 mCrab), being a
typical source of $\sim 10^{37}$ erg s$^{-1}$, at the Galactic
center distance. The source spectrum can be recovered extremly well 
even in 5 ks observation (see Fig. 3). The plasma temperature and 
optical depth can be obtained with a $90\%$ error of $\sim 5\%$.
These numbers testify for the extreme spectral accuracy that can be 
obtained and open the window of the study of spectral variability
looking directly at the physical parameters of the models. 

Furthermore, we simulate the addition of a very faint hard tail in 
the spectrum of a bright Z source. For a source flux we took $2\times 
10^{-9}$ erg cm$^{-2}$ s$^{-1}$ (i.e. about 120 mCrab) and for the tail 
we took a flux of $6\times 10^{-12}$ erg cm$^{-2}$ s$^{-1}$, i.e. $0.1\%$ 
of the main flux. Even at this faint level, the tail can be recovered 
in a 40 ks observation. Observations of Z sources can therefore shed light 
on the presence of these tails and on their evolution as the source 
moves along the color-color diagram.

\begin{figure}[]
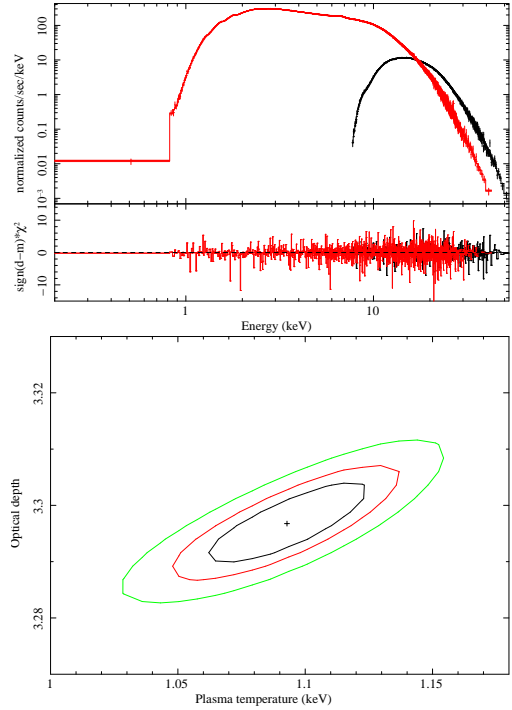

\resizebox{\hsize}{!}{\includegraphics[clip=true,angle=-90]{campana_f4.ps}}
\resizebox{\hsize}{!}{\includegraphics[clip=true,angle=-90]{campana_f5.ps}}
\caption{\footnotesize Simulated spectrum of a typical atoll source as observed 
with Simbol-X. The upper figure shows the simulated spectrum for 5 ks exposure 
and for a source flux of $10^{-9}$ erg cm$^{-2}$ s$^{-1}$ (0.5--20 keV). The 
lower figure shows the countour plot of the plasma temperature and the optical depth 
of the best fit spectrum, highlighting the very small errors on these parameters.}
\label{cont}
\end{figure}

\subsection{Study of hard tails in quiescent neutron star sources}

In order to exploit the Simbol-X capabilities we address here 
the study of the hard energy tails in quiescent LMXB transients.
These sources have received a boost of interest after Chandra 
and XMM-Newton, the first two instruments able to provide 
detailed spectral information at these low quiescent fluxes.

Recent data show that these hard tails are not present in all 
transient sources in quiescence. Their overall behaviour shows 
a hint of an anti-correlation of the hard tail luminosity 
with the total quiescent luminosity, even if more data are needed
(Jonker et al. 2004).
In addition, works on the brightest sources have shown that 
spectral changes are occurring in quiescence and these can be (also)
interpreted as variations in the hard tail (Rutledge et al. 2001; Campana
\& Stella 2003). 

We simulated the quiescent spectrum of the well known transient Aql X-1.
The source is well detected up to 80 keV in a 100 ks observation.
This will allow us to search for spectral variability on a shorter
timescale at low (as already observed in Cen X-4, Campana et al. 2004b) 
and high energies. In addition, we can gain insight on the presence
of a high energy cutoff, opening the possibility of a physical characterization
of these tails.

We also investigated the observability of the hard tail in the SAX J1808.4--3658.
The best fit quiescent spectrum comes from a deep XMM-Newton observation, confirming 
the hardness of the spectrum (photon index $\Gamma=1.8$, Heinke et al. 2007). 
In a 100 ks exposure we can have a detection up to 50 keV (as well as an extremely 
detailed spectrum at low energies and the possibility to search for X--ray 
pulsations in quiescence). As above, this extension in energy band can lead us to
a better understanding of the physics of these sources.

\begin{figure}[]
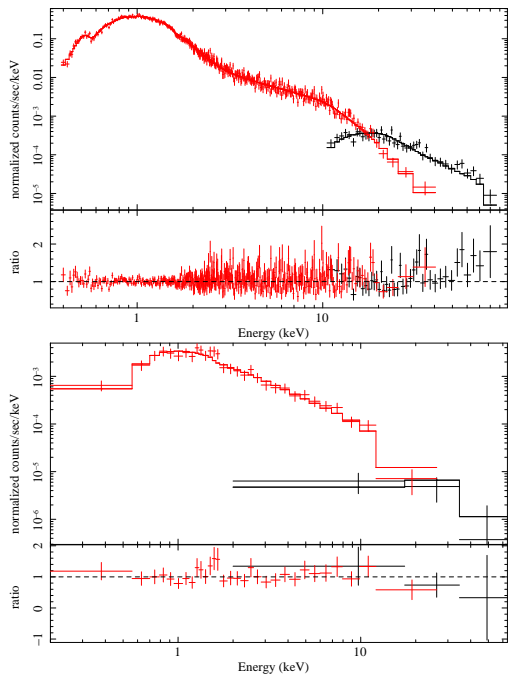

\resizebox{\hsize}{!}{\includegraphics[clip=true,angle=-90]{campana_f6.ps}}
\resizebox{\hsize}{!}{\includegraphics[clip=true,angle=-90]{campana_f7.ps}}
\caption{\footnotesize Upper panel: simulated spectrum of Aql X-1 
in quiescence as observed by Simbol-X with 100 ks exposure.
Lower panel: simulated spectrum of SAX J1808.4--3658 as observed 
by Simbol-X in 100 ks assuming as template the XMM-Newton spectrum. 
}
\label{aql1808}
\end{figure}

\section{Conclusions}

Simbol-X is the first observatory able to disclose the high energy part of 
the spectrum at very high S/N and for persistent and transient sources, 
making possible a thorough study of the emission mechanisms through 
monitoring the variability.

Different mechanisms for explaining the transition to and the quiescent 
emission of NS X--ray transient have been proposed. Simbol-X observation
will allow us to approach the problem from the point of view of physics.


\bibliographystyle{aa}

\end{document}